\documentstyle[amssymb,aps,prl,psfig,floats]{revtex}

\begin{document}

\twocolumn[\hsize\textwidth\columnwidth\hsize\csname   
@twocolumnfalse\endcsname

\title{High sensitivity two-photon spectroscopy 
in a dark optical trap, based on electron shelving.}
\author{L. Khaykovich, N. Friedman, S. Baluschev, D. Fathi, and N. Davidson}
\address{Department of Physics of Complex Systems,
Weizmann Institute of Science, Rehovot 76100, Israel}

\maketitle

\begin{abstract}
We propose a new spectroscopic method for measuring weak transitions
in cold and trapped atoms, which exploits the long interaction times and tight 
confinement offered by dark optical traps together with an electron shelving technique
to achieve extremely high sensitivity. We demonstrate our scheme by measuring a 
$5S_{1/2}\rightarrow 5D_{5/2}$ two-photon transition in cold Rb atoms trapped in a 
new single-beam dark optical trap, using an extremely weak probe laser power of 
$25$ $\mu$W. We were able to measure transitions with as small excitation rate 
as 0.09 sec$^{-1}$.\\

PACS number(s): 39.30.+w, 32.80.Pj, 32.80.Rm, 32.90.+a\\
\end{abstract}]

The strong suppression of Doppler and time-of-flight broadenings due to the
ultra low temperatures, and the possibility to obtain very long interaction
times are obvious advantages of using cold atoms for spectroscopy.
Convincing examples of such precision spectroscopic measurements are cold
atomic clocks \cite{clocks}. For RF clock transitions long interaction time
is usually obtained in an atomic fountain \cite{fountain}, while for optical
metastable clock transitions free expanding atomic clouds are used \cite
{Ertmer98}.

Even longer interaction times can be obtained for cold atoms trapped in
optical dipole traps\cite{Chu86}. To obtain long atomic coherence times,
spontaneous scattering of photons and energy level perturbations caused by
the trapping laser are reduced by increasing the laser detuning from
resonance \cite{Heinzen93}. To further reduce scattering, blue-detuned
optical traps, where repulsive light forces confine atoms mostly in the dark
(dark traps), have been developed, achieving atomic coherence of 7 s \cite
{Nir95}. The wide use of dark traps was limited by relatively complex setups
that require multiple laser beams or gravity assistance. Recent development
of single-beam dark traps make them more attractive for precision
spectroscopy \cite{Ozeri99},\cite{opn99}.

Dark traps have an additional advantage that makes them especially useful
for the spectroscopic measurements of extremely weak optical transitions.
While preserving long atomic coherence times those traps can provide large
spring constants and tight confinement of trapped atoms \cite{Ozeri99} to
ensure good spatial overlap even with a tightly focused excitation laser
beam. Therefore the atoms can be exposed to a much higher intensity of the
excitation laser, yielding a further increase in sensitivity for very weak
transitions.

In this letter we present a new and extremely sensitive method for measuring
weak transitions with cold atoms in a far detuned single-beam dark trap
using electron shelving spectroscopy \cite{electrShelv}. Recently, a similar
technique was adapted to demonstrate quantum-limited detection of
narrow-linewidth transitions on a free expanding cold atomic cloud \cite
{Riehle98}. Our scheme is based on a $\Lambda $ system. Atoms with two
ground states (for example, two hyperfine levels) are stored in the trap in
a level $\left| g_{1}\right\rangle $ that is coupled to the upper (excited)
state, $\left| e\right\rangle $, by an extremely weak transition. An atom
that undergoes the weak transition, may be shelved by a spontaneous Raman
transition on the second ground level, $\left| g_{2}\right\rangle $, that is
uncoupled to the excited level by the weak transition. After waiting long
enough, a significant fraction of the atoms will be shelved on this second
level. Finally, the detection scheme benefits from the multiply excited
fluorescence of a strong closed transition from $\left| g_{2}\right\rangle $%
, that utilizes quantum amplification due to the electron shelving technique.

We realized this scheme on a $5S_{1/2}\rightarrow 5D_{5/2}$ two-photon
transition in cold and trapped $^{85}$Rb atoms (see Fig. 1 for the relevant
energy levels) using extremely weak ($25$ $\mu $W) laser beam and we were
able to measure transitions with an excitation rate as small as 0.09 s$^{-1}$%
.

\begin{figure}[tbp]
\center{\psfig{figure=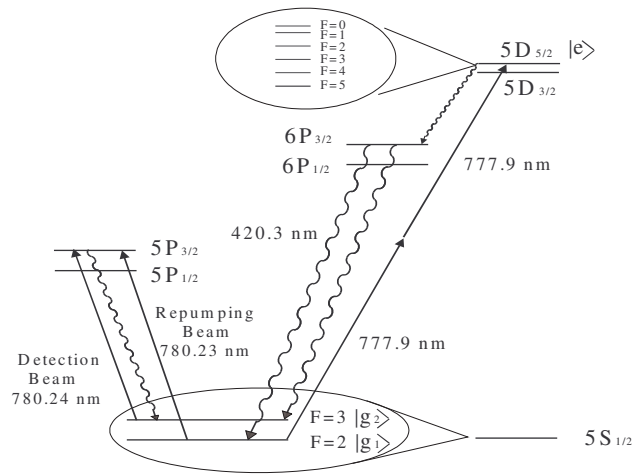,width=8.6cm}}
\caption{Energy levels of $^{85}$Rb and the transitions between them which
are involved in the experiment. Spectroscopy of the $\left|
g_{1}\right\rangle \rightarrow \left| e\right\rangle $ transition ($%
5S_{1/2}F=2\rightarrow 5D_{5/2}F^{\prime }$ in the case of $^{85}$Rb) is
performed. Atoms which undergo the transition are shelved in the level $%
\left| g_{2}\right\rangle $ ($5S_{1/2}F=3$ in $^{85}$Rb), from which they
are detected using a cycling transition (to $5P_{3/2}F=4$).}
\end{figure}

Precision spectroscopy of the two-photon transition in Rb atoms was
previously demonstrated in a hot vapor with much higher laser power \cite
{Zapka83}\cite{Nez93}. In cold Rb atoms this transition was measured either
on free expanding atoms using a mode-locked laser \cite{RiisModlock}\cite
{RiisFM} or on atoms trapped in a doughnut mode magneto-optical trap\cite
{RiisDoughnut}. In all those schemes the fluorescent $420$ nm photons were
used to detect the two-photon transition.

Our spectroscopic measurement was made on cold $^{85}$Rb atoms trapped in a
rotating-beam optical trap (ROBOT). The operation principles of the ROBOT
are described elsewhere \cite{ROBOT}. Briefly, a linearly polarized, tightly
focused ($16$ $\mu $m $1/e^{2}$ radius) Gaussian laser beam is rapidly ($100$
kHz) rotated by two perpendicular acousto-optic scanners, as seen in Fig. 2.
This forms a dark volume which is completely surrounded by a time-averaged
dipole potential walls. The wavelength of the trapping laser was $770$ nm ($%
10$ nm above the D$_{2}$ line) and its power was $380$ mW. The initial
radius of the rotation was optimized for efficient loading of the ROBOT from
a magneto-optical trap (MOT). $700$ ms of loading, $47$ ms of compression
and $3$ ms of polarization gradient cooling produced a cloud of $\sim
3\times 10^{8}$ atoms, with a temperature of $9$ $\mu $K and a peak density
of $1.5\times 10^{11}$ cm$^{-3}$. On the last stage of the loading
procedure, the atoms were optically pumped into the $F=2$ ground state by
shutting off the repumping laser $1$ ms before shutting off the MOT beams.

\begin{figure}[tbp]
\center{\psfig{figure=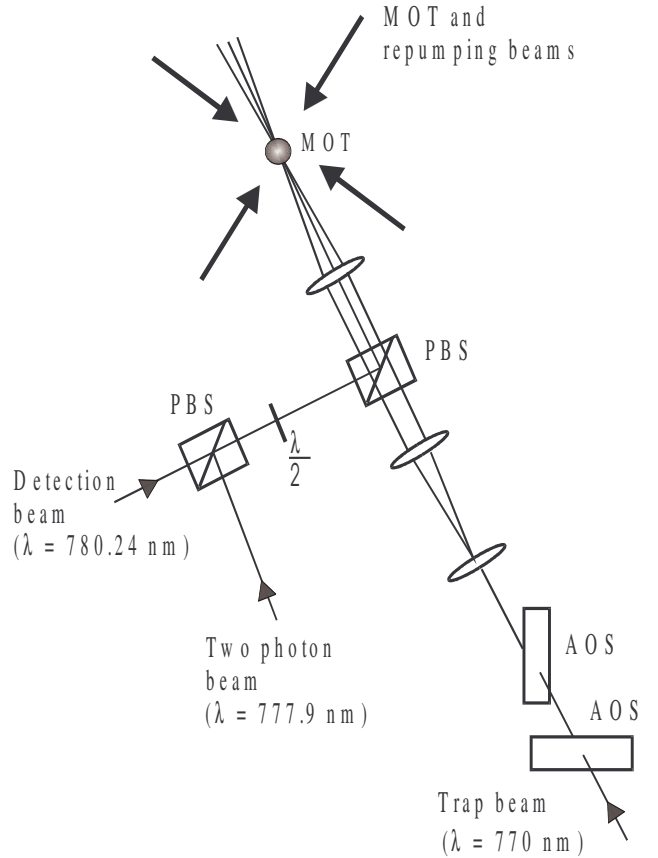,width=8.6cm}}
\caption{Schematic diagram of the experimental setup. Two acousto-optic
scanners (AOS) rotate a 10 nm blue-detuned laser beam that produce the ROBOT
trap. The two-photon beam and the detection beam are co-aligned with the
elongated axis of the trap.}
\end{figure}

After all laser beams were shut off (except for the ROBOT beam which was
overlapping the center of the MOT), $\sim 3\cdot 10^{5}$ atoms were
typically loaded into the trap, with temperature and density comparable with
those of the MOT. Next, we adiabatically compressed the trap by reducing the
radius of rotation of the trapping beam from $70$ $\mu $m to $29$ $\mu $m
such that the atoms will match the waist of the two-photon laser, to further
increase the efficiency of the transition. The size of the final cloud in
the radial direction was measured by absorption imaging and the temperature
of the atoms was measured by time of flight fluorescence imaging. From these
measurements and using our precise characterization of the trapping
potential \cite{ROBOT}, the parameters of the final cloud are: radial size ($%
1/e^{2\text{ }}$radius) of $19$ $\mu $m, axial size of $750$ $\mu $m, rms
temperatures of $55$ $\mu $K [$9$ $\mu $K] in the radial [axial] direction,
and a density of $7\cdot 10^{11}$atoms/cm$^{-3}$. The $1/e$ lifetime of
atoms in the trap was measured to be $350$ ms for both hyperfine
ground-states and was limited by collisions with background atoms. We
measured the spin relaxation time of the trapped atoms to be $>1$ s, by
measuring spontaneous Raman scattering between the two ground state levels 
\cite{Heinzen94}\cite{Ozeri99}.

The spectroscopy was performed with an external-cavity diode laser which was
tuned to the $5S_{1/2}F=3\rightarrow 5D_{5/2}F^{\prime }$ two-photon
transition ($777.9$ nm) and was split into two parts. The first part ($10$
mW) was used to frequency stabilize the laser using the $420$ nm
fluorescence signal from the two-photon excitation obtained from a $130^{0}C$
Rb vapor cell. The laser was focused into the cell to $\sim 100$ $\mu $m $%
1/e^{2}$ radius and reflected back to obtain Doppler-free spectra. We locked
the laser to the atomic line either by Zeeman modulation technique\cite
{zeeman} or directly to the side of the line. From the locking signal we
estimated the peak-to-peak frequency noise of the laser to be $\sim 3\ $MHz.
The second part of the diode laser beam passed through an acousto-optic
modulator that shifted the laser frequency toward two-photon resonance with
the $5S_{1/2}F=2\rightarrow 5D_{5/2}F^{\prime }$ transition. The laser beam
was then focused to a $26$ $\mu m$ ($1/e^{2}$ radius) spot size in the
center of the vacuum chamber, in order to optimize the efficiency of the
two-photon transition and was carefully aligned with the long (axial) axis
of the ROBOT.

We used a normalized detection scheme to measure the fraction of atoms
transferred to $F=3$ by the two-photon laser. To detect the total number of
atoms in the trap we applied a strong $200$ $\mu $s laser pulse, resonant
with the $5S_{1/2}F=3\rightarrow 5P_{3/2}F=4$ closed transition together
with the repumping laser and imaged the fluorescent signal on
photomultiplier tube (PMT). To measure only the $F=3$ population we applied
the detection pulse without the repumping laser. The $F=3$ atoms were
simultaneously accelerated and Doppler-shifted from resonance by the
radiation pressure of the detection beam within the first $100$ $\mu $s of
the pulse. Then we could detect the $F=2$ atoms by switching on the
repumping laser that pumped $F=2$ population to the $F=3$ state where atoms
were measured by the second part of the detection pulse. This normalized
detection scheme is insensitive to shot-to-shot fluctuations in atom number
as well as fluctuations of the detection laser frequency and intensity.

After the adiabatic compression of the atoms in the ROBOT was completed, the
two-photon laser on resonance with $5S_{1/2}F=2\rightarrow 5D_{5/2}F=4$ was
applied for various time intervals and the resulting $F=3$ normalized
population fraction was detected. The results for a $170$ $\mu $W two-photon
laser are presented on Fig. 3. After $100$ ms, $\sim 85\%$ of the atoms are
pumped to the $F=3$ state. This steady state population is less then 100\%
since spontaneous Raman scattering from the trapping laser and from the
two-photon laser (absorption of {\it one} photon followed by spontaneous
emission ) tend to equalize the populations of the two ground levels and
therefore compete with the measured two-photon process. The characteristic $%
1/e$ time of the four-photon spontaneous Raman scattering process which is
induced by the two-photon laser ($5S_{1/2}F=2\rightarrow 5D_{5/2}F^{\prime
}\rightarrow 6P_{3/2}F^{\prime }\rightarrow 5S_{1/2}F=3$, see Fig. 1) is
obtained from a fit to the data as $\tau _{4p}=25$ ms. The corresponding
(four-photon) rate is $\gamma _{4p}=1/\tau _{4p}=40$ s$^{-1}$. Using the
theoretical value of the two-photon cross-section of $\sigma =0.57\times
10^{-18}\ $cm$^{4}$/W \cite{Marinescu94}, the exact branching ratio ($68\%$)
for the two-photon excitation to decay to $F=3$ \cite{sobelman}, and our
maximal excitation laser intensity of $16$ W/cm$^{2}$ we calculate $\gamma
_{4p}=391\ $s$^{-1}$, a factor of $\sim 10$ larger than the measured rate.
Using a measured value for the two-photon cross-section \cite{Collins93}
yields a somewhat larger value of $\gamma _{4p}=823\ $s$^{-1}$.

\begin{figure}[tbp]
\center{\psfig{figure=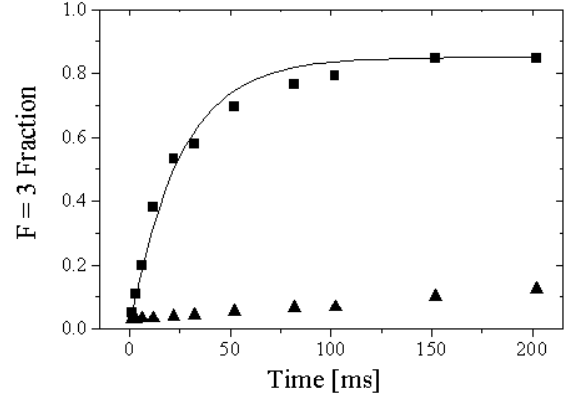,width=8.6cm}}
\caption{$F=3$ normalized population fraction as a function of the
interrogation time of the $170$ $\mu $W two-photon laser tuned to resonance
with the $5S_{1/2}F=2\rightarrow 5D_{5/2}F=4$ line ($\blacksquare $). The
solid line is a fit of the measurements by the function $%
N_{F=3}/N_{total}=A(1-e^{-t/\tau _{4p}})$, resulting $A=0.85$ as the steady
state population, and $\tau _{4p}=25$ ms as the four-photon spontaneous
Raman scattering time (see text). Spontaneous Raman scattering rate caused
by trapping laser is also given ($\blacktriangle $).}
\end{figure}

The main factor that reduced the measured excitation rate was the linewidth
of the two-photon laser that was $\sim 6$ times larger than the $300$ kHz
natural linewidth of the two-photon transition\cite{Nez93}. The
inhomogeneous broadening due to Stark-shift was calculated for the
compressed trapping potential to be $\sim 400$ kHz, which is smaller than
the laser linewidth , hence it does not contribute to the reduced excitation
rate. An additional reduction of the excitation rate may be caused by
imperfect matching between the trapped atomic sample and the maximal
intensity of the two-photon laser, so the overall agreement between the
measured and the expected $\gamma _{4p}$ is reasonable.

To measure the excitation spectrum of the $5S_{1/2}F=2\rightarrow
5D_{5/2}F^{\prime }$ transition we scanned the frequency of the two-photon
laser using the acousto-optic modulator. For each frequency point the whole
experimental cycle was repeated, with $50$ ms interrogation time of the
two-photon laser. The $F=3$ fraction of atoms as a function of the frequency
of the two-photon laser is presented in Fig. 4a. A $1.75$ MHz linewidth
(FWHM) of the atomic lines was determined by fitting the data with a
multi-peak Gaussian function and is limited by the linewidth of the
two-photon laser. This measurement agrees well with the frequency noise of
the laser estimated from the locking signal. The distances between the lines
obtained from this fit are $4.48$ MHz, $3.76$ MHz and $2.76$ MHz, and are in
excellent agreement with previously reported values of $4.50$ MHz, $3.79$
MHz and $2.74$ MHz \cite{Nez93}. The height-ratio between the lines obtained
from the fit are $1:0.86:0.47:0.21$ for $F^{\prime }=4,3,2,1$ respectively.
The expected values were calculated using the strength of the two-photon
transitions \cite{Nez93} together with the two photon decay via the $6P_{3/2}
$ level\cite{sobelman} to be $1:0.85:0.4:0.1$, in good agreement with the
measured values, except for the weakest line. Note that although the
two-photon transition $F=2\rightarrow F^{\prime }=0$ is allowed, a
two-photon decay with $\Delta F=3$ is forbidden and therefore this line is
not detected.

\begin{figure}[tbp]
\center{\psfig{figure=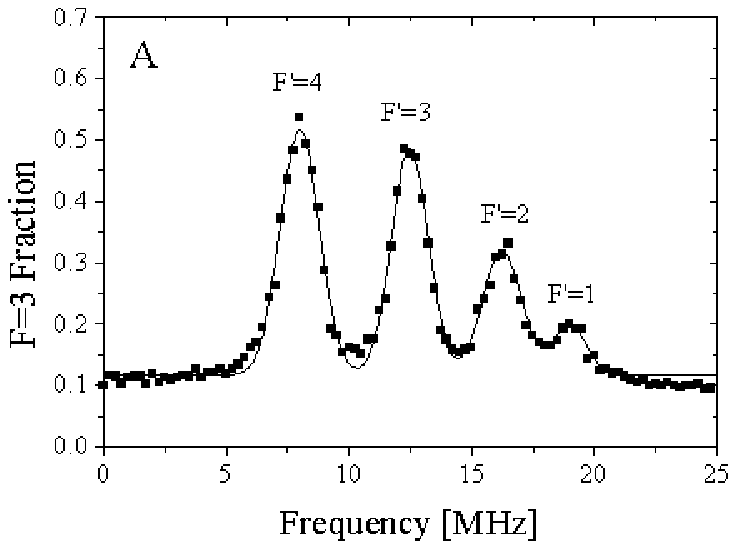,width=8.6cm}}
\center{\psfig{figure=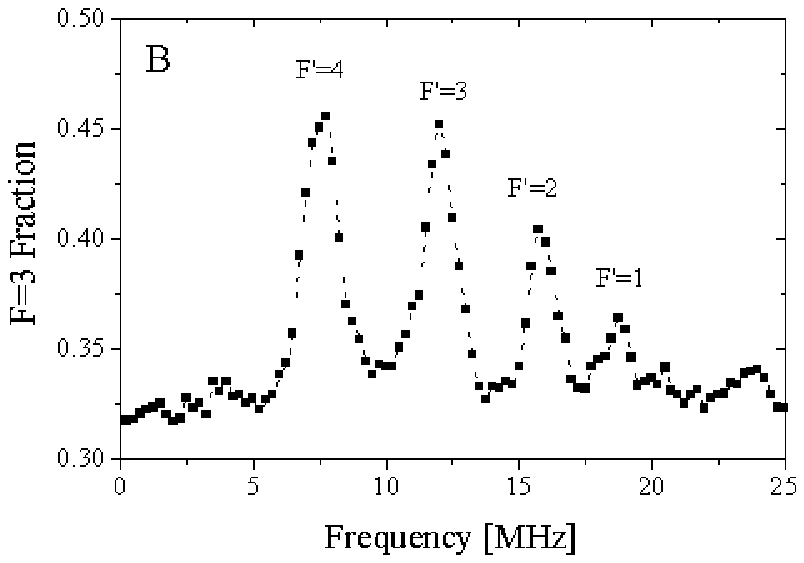,width=8.6cm}}
\caption{
A: Frequency scan of the $5S_{1/2}F=2\rightarrow 5D_{5/2}F^{\prime
}=4,3,2,1$ line of the two-photon transition, after $50$ ms exposure to a $%
170$ $\mu $W two-photon laser. The solid line is a fit to the data by a
multi-peak Gaussian function (see text). B: The same frequency scan as in
(A), after $500$ ms exposure to a $25$ $\mu $W two-photon laser. (The dashed
curve connects the points and is given to guide the eye).}
\end{figure}

Finally, we reduced the power of the two-photon laser to $25$ $\mu $W, which
reduced the transition rate by a factor of $46$. Here, the interrogation
time of the two-photon laser was $500$ ms and the measured $F=3$ population
is shown in Fig. 4b. A spectrum similar to that taken with higher intensity
is observed. A transition rate as small as $0.09$ s$^{-1}$ (for the $%
F=3\rightarrow F^{\prime }=1$ transition) is detected in this scan. The
''quantum rate amplification'' due to electron shelving (the ratio between
the measured $\gamma _{4p}$ transition rate and the rate of the one-photon
transition used for detecting the F=3 population) is $\sim 10^{7}$ for this
case.

In conclusion, we demonstrate a new and extremely sensitive scheme to
measure weak transitions using cold atoms. The key issues in our scheme are
the long spin relaxation times combined with tight confinement of the atoms
in a dark optical dipole trap, and the use of a shelving technique to
enhance the signal to noise ratio. We demonstrated our scheme by measuring a
two-photon transition $5S_{1/2}\rightarrow 5D_{5/2}$ for $^{85}$Rb atoms
trapped in a far-detuned rotating beam dark trap using only $25$ $\mu $W
laser power. The huge quantum amplification due to electron shelving
increases the sensitivity of our scheme far beyond the photon shot noise and
technical noise encountered in the direct detection of two-photon induced
fluorescence \cite{Zapka83}\cite{Nez93}\cite{RiisModlock} \cite{Collins93}.

Our measurements may be improved in several ways. Improvements of the
lifetime and spin relaxation time of atoms in the trap will allow much
longer observation times and enable detection of much weaker transitions.
This can be done by increasing the trapping laser detuning, where even
longer spin-relaxation times are expected due to quantum interference
between the two D lines\cite{Heinzen94}. Reduction of the linewidth of the
two-photon laser will allow further improvements in the sensitivity of our
scheme. It can also be combined with mode-locked laser spectroscopy \cite
{RiisModlock} to obtain even larger sensitivities for a given time-average
power of the laser. Finally, our technique can be applied for other weak
(forbidden) transitions such as optical clock transitions\cite{Ertmer98}\cite
{Riehle98} and parity violating transitions where a much lower mixing with
an allowed transition could be used.

\end{document}